\documentclass[a4paper,11pt]{article}
\usepackage{pos}
\usepackage{xcolor}
\usepackage{acronym}

\acrodef{bbh}[BBH]{binary black hole}
\acrodef{nsbh}[NSBH]{neutron star--black hole}
\acrodef{bns}[BNS]{binary neutron star}
\acrodef{lvk}[LVK]{LIGO-Virgo-KAGRA}
\acrodef{em}[EM]{electromagnetic}
\acrodef{gw}[GW]{gravitational wave}
\acrodef{grb}[GRB]{gamma-ray burst}
\acrodef{tess}[TESS]{the Transiting Exoplanet Survey Satellite}
\acrodef{winter}[WINTER]{the Wide-field Infrared Transient ExploreR}
\acrodef{fov}[FOV]{field-of-view}
\acrodef{em1}[EM1]{Extended Mission 1}
\acrodef{em2}[EM2]{Extended Mission 2}
\acrodef{ffi}[FFI]{full-frame image}
\acrodef{agn}[AGN]{active galactic nucleus}
\acrodef{nir}[NIR]{near-infrared}
\acrodef{snr}[S/N]{signal-to-noise ratio}
\acrodef{sed}[SED]{spectral energy distribution}
\acrodefplural{sed}[SEDs]{spectral energy distributions}
\acrodefplural{fov}[FOVs]{fields-of-view}

\renewcommand{\logo}{\relax}

\title{Multi-messenger astrophysics in the gravitational-wave era}

\author*[a,b]{Geoffrey Mo}
\author[b]{Rahul Jayaraman}
\author[b]{Danielle Frostig}
\author[b]{Michael M. Fausnaugh}
\author[a,b]{Erik Katsavounidis}
\author[b]{George R. Ricker}

\affiliation[a]{
MIT LIGO Lab \& MIT Kavli Institute for Astrophysics and Space Research,\\ Massachusetts Institute of Technology, Cambridge, MA, USA}
\affiliation[b]{
MIT Kavli Institute for Astrophysics and Space Research, Massachusetts Institute of Technology, Cambridge, MA, USA}

\emailAdd{gmo@mit.edu}
\emailAdd{rjayaram@mit.edu}
\emailAdd{frostig@mit.edu}
\emailAdd{Michael.Fausnaugh@ttu.edu}
\emailAdd{kats@mit.edu}
\emailAdd{grr@mit.edu}

\abstract{
The observation of GW170817, the first binary neutron star merger observed in both gravitational waves (GW) and electromagnetic (EM) waves, kickstarted the age of multi-messenger GW astronomy. 
This new technique presents an observationally rich way to probe extreme astrophysical processes.
With the onset of the LIGO--Virgo--KAGRA Collaboration's O4 observing run and wide-field EM instruments well-suited for transient searches, multi-messenger astrophysics has never been so promising. 
We review recent searches and results for multi-messenger counterparts to GW events, and describe existing and upcoming EM follow-up facilities, with a particular focus on WINTER, a new near-infrared survey telescope, and TESS, an exoplanet survey space telescope.
}

\FullConference{XVIII International Conference on Topics in Astroparticle and Underground Physics (TAUP2023)\\
 28.08-01.09.2023\\
University of Vienna\\}

\begin{document}
\maketitle

\section{Introduction}
\vspace{-0.2cm}
Multi-messenger astrophysics with \acp{gw} is a young but rapidly growing field, spurred on by \ac{gw} detectors such as the \ac{lvk} network \citep{TheLIGOScientific:2014jea, TheVirgo:2014hva, KAGRA:2020agh} and a variety of \ac{em} follow-up instruments.
Many possible sources of multi-messenger emission exist, but the most promising are compact binary coalescences involving neutron stars, such as \ac{nsbh} and \ac{bns} mergers.
These mergers produce a wide range of \ac{em} phenomena headlined by the kilonova, a thermal transient powered by the radioactive decay of heavy elements synthesized in the merger.
We discuss efforts to search for multi-messenger emission and highlight two follow-up facilities: \ac{winter}, a new infrared survey telescope, and \ac{tess}, a space telescope primarily designed for the discovery of exoplanets.

\section{GW170817, O3, and O4}
\vspace{-0.2cm}
The discovery of the \ac{bns} merger GW170817, its companion \ac{grb}, and resulting kilonova during the LIGO--Virgo O2 observing run kickstarted \ac{gw} multi-messenger astrophysics.
This single event has led to important advances in cosmology, dense matter equation of state physics, tests of general relativity, $r$-process nucleosynthesis, and neutron star astrophysics \citep[and references within]{LIGOScientific:2017vwq, GBM:2017lvd}.
The subsequent O3 observing run was successful, with the discovery of the first \ac{nsbh} mergers and the detection of many \ac{bbh} mergers, as well as upper limits on multi-messenger \ac{gw} emission from sources such as \acp{grb} \citep{LIGOScientific:2020lst} and fast radio bursts \citep{LIGOScientific:2022jpr}.
However, no new multi-messenger sources were discovered, and the uncertainty in the \ac{bns} merger rate (representing the most promising source of \ac{gw} multi-messenger events) continues to span orders of magnitude \citep{Mandel:2021smh}.
The current \ac{lvk} observing run, O4, has thus far also not yielded any multi-messenger events, but the upcoming addition of Virgo to the detector network will improve its sensitivity and localization prospects.

\section{Current and upcoming follow-up facilities}
\vspace{-0.2cm}
Multi-messenger astrophysics is by its very nature a synergistic effort.
In addition to the \ac{lvk} \ac{gw} detectors, \ac{em} follow-up instruments across the wavelength spectrum are an integral part of multi-messenger science.
Optical all-sky surveys such as ZTF and ATLAS and dedicated follow-up networks like GRANDMA and GOTO are joined by X-ray and gamma-ray telescopes such as Fermi and Swift as well as radio arrays in searching for \ac{em} counterparts to \ac{gw} events \cite[and references within]{Antier:2020nuy, Gompertz:2020cur, Engel:2022yig}.
In the coming years, new instruments such as the gamma-ray and X-ray telescope SVOM \citep{Wei:2016eox} and the Vera Rubin Observatory \citep{LSST:2008ijt} will add to the \ac{gw} follow-up landscape.
We will focus here on introducing two instruments to the multi-messenger realm: \ac{winter} and \ac{tess}.

\subsection{WINTER}
\ac{winter} is a newly commissioned 1 sq. deg. infrared survey instrument mounted on a 1-meter telescope at Palomar Observatory \citep{Frostig:2020}.
It observes in the Y, J, and Hs (shortened H, ending at 1.7 $\mu$m) bands and has a current limiting magnitude of $\sim 18.5$~mag in the J band, with planned sensitivity upgrades.
One of its main science objectives is the discovery of kilonovae through the follow-up of \ac{gw} events.
The $r$-process elements synthesized in \ac{bns} mergers have dense line transitions at optical wavelengths, pushing the peak of the kilonova emission into the \ac{nir}.
Similarly, compared to in the optical, the \ac{nir} emission is expected to be longer-lived, and to depend less on the viewing angle of the kilonova \citep{Kasen:2017sxr}.
\ac{winter}'s \ac{nir} sensitivity combined with its relatively wide 1 sq. deg. \ac{fov} fill a unique niche in the parameter space of follow-up instruments.
A simulation study of a \ac{winter} follow-up campaign of \ac{lvk} \ac{bns} mergers using design sensitivities for \ac{winter} and \ac{lvk} O4 finds that \ac{winter} could discover up to ten kilonovae per year under ideal circumstances \citep{Frostig:2021vkt}.
Despite both the \ac{lvk} network and \ac{winter} currently falling short of their projected sensitivities, \ac{winter}'s \ac{nir} capabilities and its \ac{fov} are why it remains a powerful tool for multi-messenger discovery and characterization.

\subsection{TESS}

\begin{figure}
    \centering
    \includegraphics[width=0.85\linewidth]{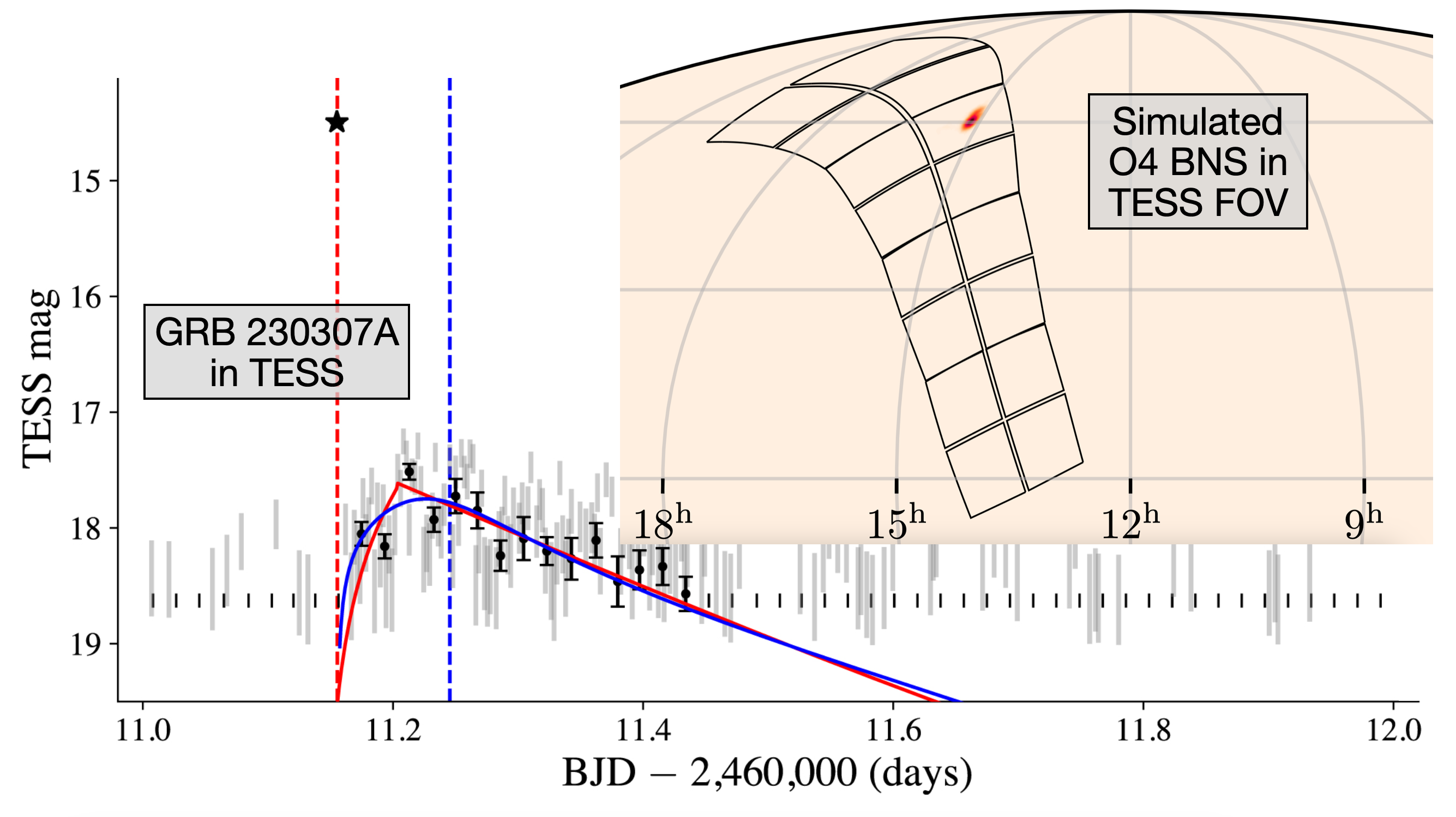}
    \caption{
Example scenario for a \ac{tess} detection of a kilonova resulting from a \ac{bns} merger discovered in \acp{gw}.
The lightcurve shows the \ac{tess} observation of GRB 230307A, including the prompt optical emission coincident with the \ac{grb} (dashed red line) and the subsequent rise and decay of the \ac{grb} afterglow.
The inset shows a simulated \ac{lvk} localization of a \ac{bns} merger using projected O4 \ac{gw} sensitivity curves (as described in \citep{Mo:2023sfz}), overlaid with the \acp{fov} of the 16 \ac{tess} CCDs.
The entire \ac{gw} localization area is contained within one \ac{tess} CCD.
If a resulting kilonova lightcurve is similarly luminous to that from GW170817, \ac{tess} will sample it as finely as it did for GRB 230307A's afterglow.
Figure modified from \citep{Mo:2023sfz, Fausnaugh:2023pmx}.
}
    \label{fig:tess}
\end{figure}

Another promising follow-up facility is \ac{tess}, which was originally designed for the discovery of exoplanets around M-dwarf stars. With its 2304 sq. deg. \ac{fov}, it has proven to be valuable for transient science as well \citep{tess:ricker}.
\ac{tess} observes its \ac{fov} for approximately one month at a time with a fixed schedule, taking an image every 200~s in a single filter spanning from 600~nm to 1000~nm. 
Each 200~s integration has a limiting magnitude of about 17.5 in the \ac{tess} band, but stacking exposures can increase \ac{tess}'s sensitivity to 20.5~mag for an 8~hour stack.
Due to its large \ac{fov} (equivalent to about 5\% of the sky), \ac{tess} is likely to make serendipitous observations of \ac{gw} skymaps, and, if it is bright enough, any counterpart optical/\ac{nir} emission.

Simulations of \ac{bns} mergers and their resulting kilonova lightcurves have shown that \ac{tess} can detect up to two kilonovae per year from \ac{bns} found in \acp{gw}. 
The simulations also show the importance of performing searches for kilonovae in \ac{tess} data, even without a \ac{gw} event: as many as eight \ac{bns} per year may be too quiet in \acp{gw} to result in an \ac{lvk} trigger, but can produce kilonovae bright enough to be detectable in \ac{tess} \citep{Mo:2023sfz}.
If a GW170817-like \ac{bns} occurs within its \ac{fov}, \ac{tess} will provide exquisite photometric measurements of the kilonova.
Fig.~\ref{fig:tess} shows this example scenario using a well-localized simulated O4 \ac{bns} and the \ac{tess} observation of the afterglow of GRB 230307A as an example light curve \citep{Fausnaugh:2023pmx}.
The above predictions depend strongly on \ac{bns} merger rates, which are a large source of uncertainty in simulations like these.

\section{Conclusion}
\vspace{-0.2cm}
As O4 continues and follow-up instruments across the \ac{em} spectrum such as \ac{winter} and \ac{tess} realize their potential, prospects for multi-messenger observations remain bright.
In addition to the \ac{bns} mergers we know can result in \ac{em} counterparts, there is also the possibility of unveiling new multi-messenger science from \ac{nsbh} and \ac{bbh} mergers, as well as potentially from other sources of \acp{gw} such as core-collapse supernovae or magnetars.
Beyond O4, instruments such as LSST on the Vera Rubin Observatory promise to revolutionize time-domain astronomy as a whole, driving forward the horizon of multi-messenger astrophysics.

\bibliographystyle{JHEP}
\bibliography{taup}

\end{document}